\pdfoutput=1

\documentclass[11pt]{article}

 \usepackage{emnlp2021}

\usepackage{times}
\usepackage{latexsym}

\usepackage{amsthm}
\usepackage{amsmath}
\usepackage{amssymb}

\usepackage[T1]{fontenc}

\usepackage[utf8]{inputenc}

\usepackage{microtype}
\newtheorem{definition}{Definition}

\usepackage{graphicx}
\usepackage{subfigure}
\usepackage{listings}

\title{Tea: Program Repair Using Neural Network Based on Program Information Attention Matrix}

\author{Wenshuo Wang\\
	Microsoft Asia-Pacific Technology Company Ltd. \\
	University of Chinese Academy of Sciences \\
	TCA, Institute of Software, Chinese Academy of Sciences \\
	\texttt{wangwenshuo18@mails.ucas.ac.cn}\\
  \AND
  Chen Wu \\
  Microsoft Asia-Pacific Technology Company Ltd. \\
  \texttt{wu.chen@microsoft.com} \\
  Liang Cheng \\
University of Chinese Academy of Sciences \\
TCA, Institute of Software, Chinese Academy of Sciences \\
  \texttt{chengliang@iscas.ac.cn} \\
  Yang Zhang \\
University of Chinese Academy of Sciences \\
TCA, Institute of Software, Chinese Academy of Sciences \\
  \texttt{zhangyang@iscas.ac.cn} \\
  }

\begin{document}
\maketitle
\begin{abstract}

The advance in machine learning (ML)-driven natural language process (NLP) points a promising direction for automatic bug fixing for software programs, as fixing a buggy program can be transformed to a translation task.  While software programs contain much richer information than one-dimensional natural language documents, pioneering work on using ML-driven NLP techniques for automatic program repair only considered a limited set of such information. We hypothesize that more comprehensive information of software programs, if appropriately utilized, can improve the effectiveness of ML-driven NLP approaches in repairing software programs. As the first step towards proving this hypothesis, we propose a unified representation to capture the syntax, data flow, and control flow aspects of software programs, and devise a method to use such a representation to guide the transformer model from NLP in better understanding and fixing buggy programs. Our preliminary experiment confirms that the more comprehensive information of software programs used, the better ML-driven NLP techniques can perform in fixing bugs in these programs.

\end{abstract}
\section{Introduction}
Finding and fixing bugs accounts for a significant portion of maintenance cost for software~\cite{britton2013reversible}, and the cost of bug fixing increases exponentially with time~\cite{dawson2010integrating}. Hence, automatic program repair has long been a focus of software engineering, with the goal of lowering down the cost and reducing the introduction of new bugs during bug fixing. 


Traditional automatic program repair approaches utilize certain aspects of information of a program to repair its bugs. For example, GenProg~\cite{genprog} used genetic algorithms to mutate the abstract syntax tree (AST) of a C program for bug fixing, while Nopol~\cite{nopol} used trace and data types collected from the target programs to repair assertion condition errors and assertion statement missing errors. This type of approaches, however, suffered low accuracy in bug detection and repair due to the lack of a good code generation model. 

Recent advance in natural language processing (NLP), especially machine learning (ML) based NLP technologies, has inspired automatic program repair research in that ML techniques designed for NLP tasks such as automated translation have been revised to improve the understanding of buggy programs and/or to derive correct program repairs that achieve that greatest repairing goals. For example,  ACS~\cite{acs} sorts variables through the principle of locality and then uses NLP technologies to analyze the content of open-source programs to generate correct patches. CoCoNut~\cite{coconut} used the neural network translation model to generate correct code based on the contextual information (e.g., the adjacent code) of bugs. 

However, these ML-based approaches only consider and utilize a limited set of information of the target program. For example, neither data flow nor control flow was considered by CoCoNut, even though it suggested that the contextual information used in its model can be replaced by any other forms of contextual information. This has limited the applicability, accuracy, and effectiveness of program repair that these technique can achieve. 

In addition, there are distinct differences between software programs and natural language documents: entities in software programs (e.g., variables, statements, and functions) are related to each other not only in syntactic or semantic bindings, but also for data- and control-dependencies, which makes software program more complex than one-dimension text.  As a result, a software program can often execute in different orders (i.e., traces), most of which are different than the syntactic order of its statement.  Hence, we hypothesize that the syntax, semantics, data flow and control flow information of software programs, if appropriately utilized, can improve the accuracy and effectiveness of automatic program repair techniques based on NLP technologies. 

As the first step towards proving this hypothesis, we propose a unified representation, called Program Information Attention Matrix (PIAM), to capture the syntax, data dependency, and control dependency of software programs to assist program repair tasks.  We also develop a prototype called TEA (\underline{T}ransformer Cod\underline{e} \underline{A}ttention) that replaces the attention mechanism of the transformer model from NLP with the PIAM, so that it can quickly grasp the comprehensive relationship between the entities in a buggy program and 'translate' into a correct program in a more accurate manner. 

We have evaluated the effectiveness of TEA in bug fixing over the Tufano dataset\footnote{\url{https://github.com/micheletufano/NeuralCodeTranslator/tree/master/dataset}.}, and the evaluation results suggested that the more aspects of program information is considered and the more granular such information is, the better TEA generally performs in bug fixing. 
\section{Background: Transformer in NLP}
The transformer in NLP~\cite{attention} is a novel architecture that revolutionized machine learning based sequence-to-sequence (seq2seq) translation, since it '\emph{relies entirely on self-attention to compute representations of its input and output without using sequence-aligned RNNs or convolution}'.  

Like other seq2seq models, the transformer also incorporates an encoder and a decoder. However, in the transformer, the encoder has one layer of a multi-head attention followed by a layer of feed forward neural network. The decoder has a similar setup but with an extra masked multi-head attention layer. The encoder transforms the word embeddings of the input sequence into an intermediate vector, while the decoder translate the intermediate vector into an output sequence with the greatest probability of achieving the translation goals.

In both encoder and decoder, the attention mechanism helps to relate the positions of a target sequence to better understand it (a notion known as 'self-attention') and the multi-head mechanism helps to calculate self-attention multiple times in a parallel and independent manner.  Self-attention ensures that information from different representation subspaces at different positions of the target sequence has been attended. Lastly, the masked multi-head mechanism in the decoder helps it focus on appropriate parts of the input sequence. 


The multi-head attention of the transformer can be calculated according to the black font part of Equations~\ref{improved_attention}. In Equation~\ref{improved_attention}, Q, K, V are three vectors representing the same sequence, $d_k$ is the dimension of K, and $softmax$ is a softmax activation function to normalize the attention scores.  The multi-head mechanism is used to  concatenates different attention matrices 
to calculate attention in different spaces. 

When applied to automatic program repair, the transformer takes as an input a buggy program and 'translates' it into a correct one, during which the self-attention mechanism guides the repair with the important relationships between program entities (e.g., tokens) to improve repair accuracy.


\section{Method}
\subsection{Program Information Attention Matrix}\label{sec:PIAM}

As aforementioned, software programs have richer information than natural language text that should be collected and utilized for program repairing. In particular, the relationship between tokens in software programs are determined not only by their relative positions, but also by many other aspects of features of these programs.

In this work, we consider control flow, data flow, and syntax of software programs, because they are representative program features that have been well studied in the program analysis area (i.e., there are a vast body of techniques and tools available for collecting these features from software programs).  Apparently, other program features may also be considered depending on the needs of program repair tasks. 

\begin{figure}[h]
    \centerline{\includegraphics[scale=0.40]{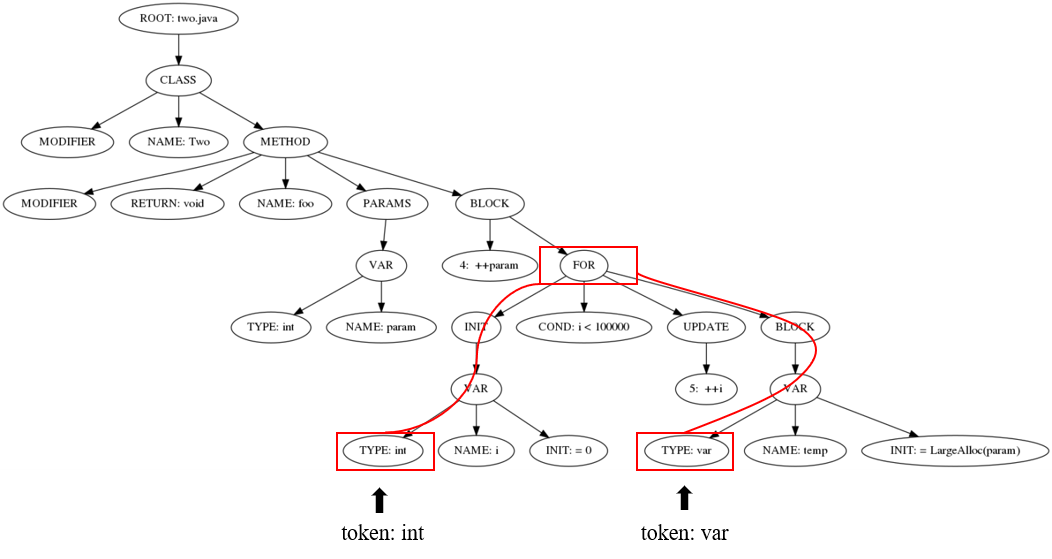}}
    \caption{Labeled AST. 
    }
    \label{fig_ast2}
\end{figure}

The syntax structure of a software program is typically captured by its AST.  For example, for the following code fragments, Figure~\ref{fig_ast2} illustrates its AST, in which leaf nodes represent statements in the code and non-leaf nodes represent the relationship between leaves. 
\begin{lstlisting}
void foo(int param)
{
    ++param;
    for (int i = 0; i < 100000; ++i)
    {
        var temp = LargeAlloc(param);
    }
}
\end{lstlisting}

Thus, the relationship between any pair of leaf nodes in an AST can be determined by:  1) their nearest common ancestor (NCA) on the AST; 2) the marked types of these two nodes and their NCAs (denoted as NCALabel); and 3) 
the nodes along the shortest path between these two nodes through it NCA (denoted as NCAPath). For example, for tokens \texttt{int} and \texttt{var} in Figure~\ref{fig_ast2}, their NCA is token \texttt{FOR}, their NCALabel feature is \texttt{TYPE->FOR->TYPE} (red circles in Figure~\ref{fig_ast2}); and their NCAPath feature is \texttt{TYPE->VAR->INIT->FOR->BLOCK-> TYPE} (the red line in Figure~\ref{fig_ast2}).

Similarly, the control flow feature of a program is captured by its CFG, which characterizes its execution paths; and its data flow feature (in other words, data dependency) is characterized by its DFG,  which reveals how variables and parameters are accessed and manipulated across the program. 
 
The AST, CFG, and DFG representations of a program each reveals a specific aspect of the program. We propose the notion of PIAM to unify these different representations of software programs, so that they can be used by the transformer model for program repair, avoiding the over-reliance on a specific representation.  More specifically, Definition~\ref{MPIAM} provides the formal definition of PIAM for software programs. 

\begin{definition}[PIAM]
Given a program $P$ that can be tokenized to a set of tokens, its PIAM is a matrix $M=x_{ijk}$ that satisfies: 
\begin{itemize}
\item $1\le i \le |tokens|$, $1\le j \le |tokens|$ and $1\le k \le |features|$, where $features$ are the set of features of $P$  considered.
\vspace{-6pt}
\item If the indexes of tokens $t1$ and $t2$ are respectively $id1$ and $id2$, and $t1$ and $t2$ are the last word and the first word of two adjacent nodes in the CFG of $P$, then $x[id1][id2][q]=1$, where $q$ is the index of the CFG feature.
\vspace{-6pt}
\item If the indexes of tokens $t3$ and $t4$ are respectively $id3$ and $id4$, and $t3$ and $t4$ are the tokens of two different leaves in $P$'s AST, then $x[id3][id4][p]=1$, where $p$ is the index of the AST feature. 
\vspace{-6pt}
\end{itemize}
\label{MPIAM}
\end{definition}

Apparently, PIAM conforms to the definition of the attention matrix of the transformer model.

In this work, we design three versions of the PIAM, each representing a different way of capturing the AST, CFG, and DFG information of software programs: 
\begin{itemize}
\item Version 1: relations considered include direct neighbor tokens in program syntax, direct adjacent nodes in the CFG and DFG, and up to x NCA features in the AST \footnote{Based on the current data sets considered, we configure x to a value between 22 and 25.} ; and $p$ in Definition~\ref{MPIAM} is set to 32. 
\item Version 2: relations considered include direct neighbor tokens in program syntax, direct adjacent nodes in the CFG and DFG, up to x NCA features in the AST, and 125-x NCALabel features of the program; and $p$ in Definition~\ref{MPIAM} is set to 128.
\item Version 3: relations considered include direct neighbor tokens in program syntax, direct adjacent nodes in the CFG and DFG, up to x NCA features in the AST, and 125-x NCAPath features of the program; and $p$ in Definition~\ref{MPIAM} is set to 128.
\end{itemize}
It is apparent that version 1, 2, and 3 of PIAM follow an increasing order in the comprehensiveness and granularity of information they consider for software programs.


\subsection{Revised Transformer Using PIAM}
We revise the attention mechanism in~\cite{attention} using the PIAM, which marked by red in  Equation~\ref{improved_attention}, in which $W$ is the coefficient for linear transformation of PIAM.

\begin{equation}
\begin{split}
Attention(Q,K,V)=&\\softmax(\frac{QK^T}{\sqrt{d_k}}{\color{red}*\alpha+W*PIAM*(1-\alpha)})V \label{improved_attention}
\end{split}
\vspace{-6pt}
\end{equation}

$\alpha$ in Equation~\ref{improved_attention} is a  parameter in the range of $[0,1]$, which can be learned through training to represent the proportional relationship between the original attention matrix and the version of PIAM used. Given an input program $x$, $\alpha$ can be calculated using Equation $\alpha=sigmod(wx+b)$, where are $w$ and $b$ are constant coefficients. 

\subsection{Implementation}
\begin{figure}[htbp]
\centerline{\includegraphics[scale=0.28]{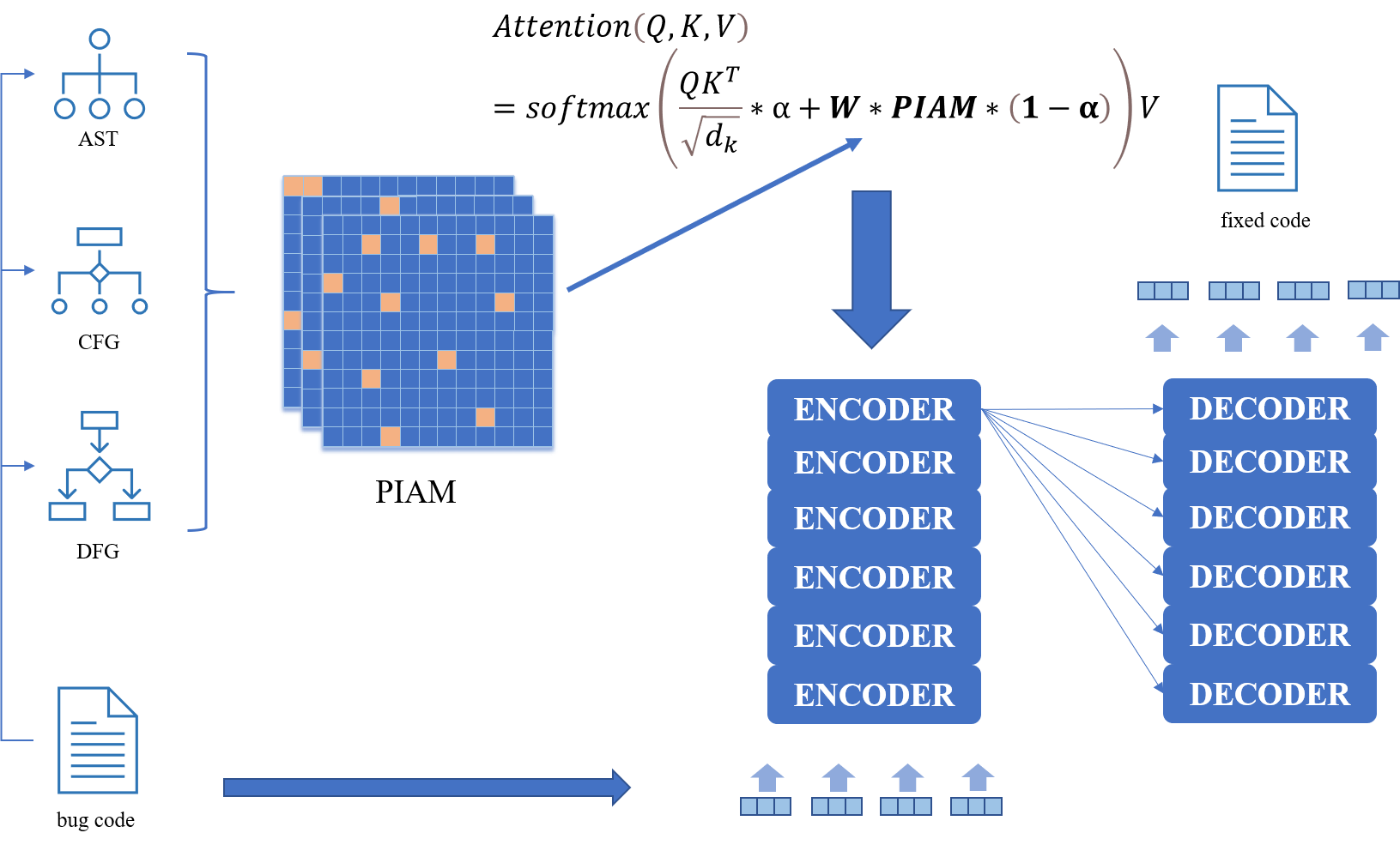}}
\caption{Architecture of the TEA Prototype.}
\label{fig:mpiam}
\end{figure}

We have implemented a prototype called TEA that utilizes the revised transformer model for bug fixing. Figure~\ref{fig:mpiam} illustrates the architecture of TEA, in which different versions of PIAM can be incorporated. 

In particular, TEA tokenizes the input program using the bpe algorithm~\cite{bpe} and utilizes traditional static analysis techniques to retrieve its AST, CFG, and DFG information. TEA revises an open-source pytorch implementation~\footnote{\url{https://github.com/SamLynnEvans/Transformer}.} of the Transformer in~\cite{attention} as the baseline transformer, and replaces its attention mechanism with one of three versions of PIAM defined in section~\ref{sec:PIAM} based on configuration. 

\section{Evaluation}

\begin{figure}[h]
	\centerline{\includegraphics[scale=0.40]{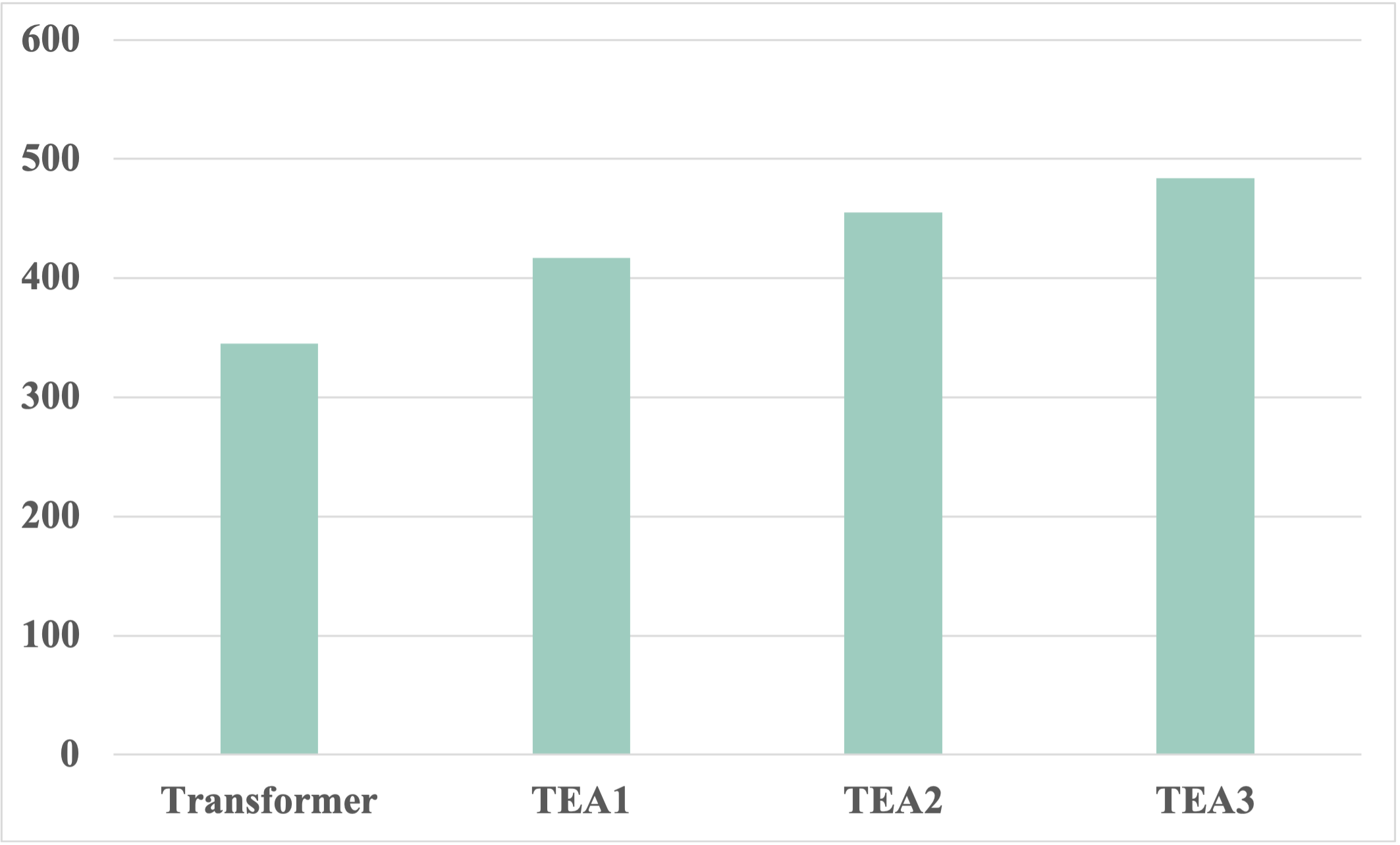}}
	\caption{Comparision with different models. 
	}
	\label{fig:sum_repair}
\end{figure}

We have evaluated the performance of TEA, with different versions of PIAM incorporated, in bug repair using a small data set called Tufano. Figure~\ref{fig:sum_repair} summarizes the evaluation results, in which Transformer, TEA1, TEA2, and TEA3 correspond to no PIAM (i.e., only the baseline transformer is used), version 1, 2, or 3 of PIAM in section~\ref{sec:PIAM} is incorporated in TEA.  It is obvious from Figure~\ref{fig:sum_repair} that these four configurations of TEA fixed more bugs (345, 417, 455, and 484 respectively) when the incorporated PIAM captures more comprehensive and granular information of the subject programs. 

\section{Related Work}

The advance in NLP, especially ML-driven NLP,  has also greatly improved the accuracy of translation tasks. This advance has influenced the automatic program repair~\cite{coconut,  chen2019sequencer}.  However, existing work on using NLP technologies to tackle automatic bug fixing tasks only considers a limited set of information of the subject programs. In contrast, our work proposes to collect and utilize different aspects of information of software programs in improve the accuracy and effectiveness of bug fixing.  


There are many forms of representation having been proposed for software programs, beside AST, CFG, and DFG considered in this work. For example, Wei Le et al. proposed a data structure called Multiversion Interprocedural Control Flow Graph (MVICFG)~\cite{mvicfg} for patch verification. This thread of work offer possibilities to expand the types of information of software programs considered by the PIAM so that it can better meet the needs of different program repair tasks.

\section{Conclusion and Future Work}
We have presented a prototype called TEA as the first step towards proving our hypothesis that utilizing different aspects of features of programs can improve the accuracy and effectiveness of the transformer model from NLP in automatically fixing bugs in software programs. Our preliminary experiment demonstrates that the more aspects of program features and the more granular of program features considered, the better the transformer generally performs in repairing program bugs. 

As next steps, we plan to train and evaluate the TEA prototype with large-scale real world program datasets, and conduct a more throughout experiment to compare its performance with other ML-driven program repair methods.

\bibliography{anthology,custom}

\begin{thebibliography}{14}
\expandafter\ifx\csname natexlab\endcsname\relax\def\natexlab#1{#1}\fi

\bibitem[{Bahdanau et~al.(2014)Bahdanau, Cho, and Bengio}]{bahdanau2014neural}
D.~Bahdanau, K.~Cho, and Y.~Bengio. 2014.
\newblock Neural machine translation by jointly learning to align and
  translate.
\newblock \emph{arXiv preprint arXiv:1409.0473}.

\bibitem[{Britton et~al.(2013)Britton, Jeng, Carver, Cheak, and
  Katzenellenbogen}]{britton2013reversible}
T.~Britton, L.~Jeng, G.~Carver, P.~Cheak, and T.~Katzenellenbogen. 2013.
\newblock Reversible debugging software.
\newblock \emph{Judge Bus. School, Univ. Cambridge, Cambridge, UK, Tech. Rep}.

\bibitem[{Chen et~al.(2019)Chen, Kommrusch, Tufano, Pouchet, Poshyvanyk, and
  Monperrus}]{chen2019sequencer}
Z.~Chen, S.~J. Kommrusch, M.~Tufano, L.~Pouchet, D.~Poshyvanyk, and
  M.~Monperrus. 2019.
\newblock Sequencer: Sequence-to-sequence learning for end-to-end program
  repair.
\newblock \emph{IEEE Transactions on Software Engineering}.

\bibitem[{Cho et~al.(2014)Cho, van Merri{\"e}nboer, Gulcehre, Bahdanau,
  Bougares, Schwenk, and Bengio}]{cho2014learning}
K.~Cho, B.~van Merri{\"e}nboer, C.~Gulcehre, D.~Bahdanau, F.~Bougares,
  H.~Schwenk, and Y.~Bengio. 2014.
\newblock Learning phrase representations using rnn encoder-decoder for
  statistical machine translation.
\newblock \emph{arXiv preprint arXiv:1406.1078}.

\bibitem[{Dawson et~al.(2010)Dawson, Burrell, Rahim, and
  Brewster}]{dawson2010integrating}
M.~Dawson, D.~N. Burrell, E.~Rahim, and S.~Brewster. 2010.
\newblock Integrating software assurance into the software development life
  cycle (sdlc).
\newblock \emph{Journal of Information Systems Technology and Planning},
  3(6):49--53.

\bibitem[{de~Souza et~al.(2018)de~Souza, Goues, and
  Camilo-Junior}]{de2018novel}
E.~F. de~Souza, C.~L. Goues, and C.~G. Camilo-Junior. 2018.
\newblock A novel fitness function for automated program repair based on source
  code checkpoints.
\newblock In \emph{Proceedings of the Genetic and Evolutionary Computation
  Conference}, pages 1443--1450.

\bibitem[{Goues et~al.(2011)Goues, Nguyen, Forrest, and Weimer}]{genprog}
C.~Le Goues, T.~Nguyen, S.~Forrest, and W.~Weimer. 2011.
\newblock Genprog: A generic method for automatic software repair.
\newblock \emph{Ieee transactions on software engineering}, 38(1):54--72.

\bibitem[{Le and Pattison(2014)}]{mvicfg}
W.~Le and S.~D. Pattison. 2014.
\newblock Patch verification via multiversion interprocedural control flow
  graphs.
\newblock In \emph{Proceedings of the 36th International Conference on Software
  Engineering}, pages 1047--1058.

\bibitem[{Lutellier et~al.(2020)Lutellier, Pham, Pang, Li, Wei, and
  Tan}]{coconut}
T.~Lutellier, H.~V. Pham, L.~Pang, Y.~Li, M.~Wei, and L.~Tan. 2020.
\newblock Coconut: Combining context-aware neural translation models using
  ensemble for program repair.
\newblock In \emph{Proceedings of the 29th ACM SIGSOFT International Symposium
  on Software Testing and Analysis}, pages 101--114.

\bibitem[{Nielebock et~al.(2020)Nielebock, Heum{\"u}ller, Kr{\"u}ger, and
  Ortmeier}]{nielebock2020using}
S.~Nielebock, R.~Heum{\"u}ller, J.~Kr{\"u}ger, and F.~Ortmeier. 2020.
\newblock Using api-embedding for api-misuse repair.
\newblock In \emph{Proceedings of the IEEE/ACM 42nd International Conference on
  Software Engineering Workshops}, pages 1--2.

\bibitem[{Sennrich et~al.(2015)Sennrich, Haddow, and Birch}]{bpe}
R.~Sennrich, B.~Haddow, and A.~Birch. 2015.
\newblock Neural machine translation of rare words with subword units.
\newblock \emph{arXiv preprint arXiv:1508.07909}.

\bibitem[{Vaswani et~al.(2017)Vaswani, Shazeer, Parmar, Uszkoreit, Jones,
  Gomez, Kaiser, and Polosukhin}]{attention}
A.~Vaswani, N.~Shazeer, N.~Parmar, J.~Uszkoreit, L.~Jones, A.~N. Gomez, \L.
  Kaiser, and L.~Polosukhin. 2017.
\newblock \href
  {https://proceedings.neurips.cc/paper/2017/file/3f5ee243547dee91fbd053c1c4a845aa-Paper.pdf}
  {Attention is all you need}.
\newblock In \emph{Advances in Neural Information Processing Systems},
  volume~30. Curran Associates, Inc.

\bibitem[{Xiong et~al.(2017)Xiong, Wang, Yan, Zhang, Han, Huang, and
  Zhang}]{acs}
Y.~Xiong, J.~Wang, R.~Yan, J.~Zhang, S.~Han, G.~Huang, and L.~Zhang. 2017.
\newblock Precise condition synthesis for program repair.
\newblock In \emph{2017 IEEE/ACM 39th International Conference on Software
  Engineering (ICSE)}, pages 416--426. IEEE.

\bibitem[{Xuan et~al.(2016)Xuan, Martinez, Demarco, Clement, Marcote, Durieux,
  Berre, and Monperrus}]{nopol}
J.~Xuan, M.~Martinez, F.~Demarco, M.~Clement, S.~L. Marcote, T.~Durieux, D.~Le
  Berre, and M.~Monperrus. 2016.
\newblock Nopol: Automatic repair of conditional statement bugs in java
  programs.
\newblock \emph{IEEE Transactions on Software Engineering}, 43(1):34--55.

\end{thebibliography}
\bibliographystyle{acl_natbib}




\end{document}